\documentclass[a4paper,11pt]{article}
\pdfoutput=1 
\usepackage{jheppub}
\usepackage{amssymb} 
\usepackage{amsmath}
\usepackage{mathtools}
\usepackage{amsfonts}    
\usepackage{dsfont}
\usepackage{young}
\usepackage[vcentermath]{youngtab}
\usepackage{bm}
\usepackage{braket}

\newcommand{\be}{ \begin{equation}}
\newcommand{\ee}{\end{equation}}
\newcommand{\bea}[1]{\begin{eqnarray}\label{#1} }
\newcommand{\eea}{\end{eqnarray}}

\def\ZZZ{{\hskip-3pt\hbox{ Z\kern-1.6mm Z}}}
\def\zzz{{\hskip-3pt\hbox{ z\kern-1mm z}}}

\def\one{{\hbox{ 1\kern-.8mm l}}}
\def\zero{{\hbox{ 0\kern-1.5mm 0}}}

\newcommand{\mrm}[1]{\mathrm{#1}}
\newcommand{\mcl}[1]{\mathcal{#1}}
\newcommand{\mbb}[1]{\mathbb{#1}}
\newcommand{\mbf}[1]{\mathbf{#1}}
\newcommand{\mfr}[1]{\mathfrak{#1}}

\title{BPS Correlators for $\text{AdS}_3/\text{CFT}_2$}

\author[]{Matthias R.\ Gaberdiel and} 
\author[]{Beat Nairz}

\affiliation[]{Institut f\"ur Theoretische Physik, ETH Zurich, \\
CH-8093 Z\"urich, Switzerland}

\emailAdd{gaberdiel@itp.phys.ethz.ch}
\emailAdd{nairzb@student.ethz.ch}

\abstract{The BPS correlators of the symmetric product orbifold $\text{Sym}_N(\mbb{T}^4)$ are reproduced from the dual worldsheet theory describing strings on $\text{AdS}_3\times {\rm S}^3\times \mbb{T}^4$ with minimal ($k=1$) NS-NS flux. More specifically, we show that the worldsheet duals of the symmetric orbifold BPS states can be identified with their lift to the covering surface, thereby making the matching of the correlators essentially manifest. We also argue that the argument can be generalised to arbitrary descendants, using suitable DDF operators on the worldsheet.}

\begin{document}

\maketitle
\flushbottom

\section{Introduction}

The $\text{AdS}_3/\text{CFT}_2$ correspondence is a useful playground to study aspects of the AdS/CFT  correspondence explicitly. In particular, string theory on  $\text{AdS}_3\times {\rm S}^3\times \mbb{T}^4$ with minimal $k=1$ units of pure NS-NS flux is exactly dual to the free symmetric product orbifold $\text{Sym}_N(\mbb{T}^4)$ in the large $N$ limit \cite{Eberhardt:2018ouy,Eberhardt:2019ywk,Dei:2020zui}, and this has allowed for detailed precision tests. For example, the correlators of the symmetric orbifold can be calculated in terms of holomorphic covering maps \cite{Lunin:2000yv,Lunin:2001pw}, and this is reproduced from the worldsheet perspective where the worldsheet correlators localise the integral over the worldsheet moduli to those configurations for which a holomorphic covering map exists \cite{Eberhardt:2019ywk, Dei:2020zui}, see also \cite{Eberhardt:2020akk,Knighton:2020kuh} for the analysis at higher worldsheet genus. Thus, the string worldsheet indeed plays the role of the covering surface in the orbifold theory, as originally suggested in \cite{Lunin:2000yv,Lunin:2001pw,Pakman:2009zz,Pakman:2009ab}.

So far, this worldsheet analysis has only been performed for the vertex operators that are dual to the twisted sector ground states of the symmetric orbifold theory,\footnote{In addition, some simple correlators of the ${\cal N}=4$ superconformal fields were analysed in \cite{Gaberdiel:2021njm}.} and it would be important to generalise it to arbitrary single particle states of the symmetric orbifold theory. Here, we start to tackle this problem by looking at another special class of orbifold correlators: the correlators of BPS states \cite{Lunin:2001pw}. We identify the corresponding worldsheet states and then apply the techniques of \cite{Dei:2020zui} to determine their correlators up to some unfixed prefactors, finding complete agreement with the explicit formulae of \cite{Lunin:2001pw}. In the process we also note that the worldsheet duals of the BPS states can be directly identified with their lift to the covering surface, thereby making the correspondence effectively manifest. As a consequence, our results also suggest how this identification should work for correlators of arbitrary descendants: the worldsheet dual of an arbitrary state $\phi$ in the symmetric orbifold can be expressed in terms of the DDF operators of \cite{Eberhardt:2019qcl,Naderi}, which in turn matches with the lift of $\phi$ to the covering surface, implying that the corresponding correlators are manifestly the same. 
\smallskip

The paper is organised as follows. In Section~\ref{sec:orbifold}, we review the salient features of the symmetric orbifold theory, and in Section~\ref{sec:worldsheet} we do the same for the dual worldsheet theory. 
We then discuss the BPS correlators of the symmetric orbifold in Section~\ref{sec:bps_correlators}, and show that they can be reproduced from the worldsheet in Section~\ref{sec:worldsheet_calculation}, see also Appendix~\ref{sec:app_calculations} for an explicit check for the case of three-point functions. The suggested generalisation to arbitrary descendant states using the DDF operators is explained in Section~\ref{sec:formal_ddf_operators}. Finally, Section~\ref{sec:conclusion} contains our conclusions.

\section{The Symmetric Orbifold}\label{sec:orbifold}

In this section we briefly review the symmetric product orbifold following \cite{Lunin:2000yv, Lunin:2001pw}. In  particular, we explain their method of calculating path-integral correlators of twisted sector fields using branched covering maps. We also identify the BPS states of the symmetric orbifold theory.  In the following we shall not attempt to give a comprehensive review, but rather introduce what will be needed below;  more details about symmetric orbifolds can e.g.\ be found in \cite{Lunin:2000yv, Lunin:2001pw, Pakman:2009zz, Roumpedakis:2018tdb}.

\subsection{The Setup}

Given a seed theory with target manifold $M$, the symmetric product orbifold theory is defined as
\begin{equation}
    \mrm{Sym}_N(M) = M^N/S_N\ .
\end{equation}
As the action of $S_N$ is not free, this target space is not a smooth manifold. The identification of the different copies via $S_N$ results in `twisted' sectors for which the fields satisfy the twisted boundary conditions,
\begin{equation}
   v^{(k)}(e^{2\pi i}x + x_0) = v^{(\varrho(k))}(x_0)\ ,\quad \varrho \in S_N\ ,
\end{equation}
where $e^{2\pi i}x$ is a short-hand for going around a small circle centered at $x=0$, and $v^{(k)}$ denotes the field $v$ in the $k$'th copy of the product theory.\footnote{Due to the identification of the copies, these boundary conditions are associated to equivalence classes of permutations (rather than individual permutations).} We may think of these twisted boundary conditions as being the consequence of a `twist field' at $x_0$. The analogue of the `single-trace' operators of SYM are the twist fields that are associated to permutations $\varrho$ consisting of a single cycle, and we shall only consider them here (since they corresponds to the single string states of the perturbative worldsheet theory). We shall denote the $w$-cycle twist operator of smallest conformal dimension  by $\sigma_w$.

\begin{figure}[ht]
        \centering
        \includegraphics[width = .48\textwidth]{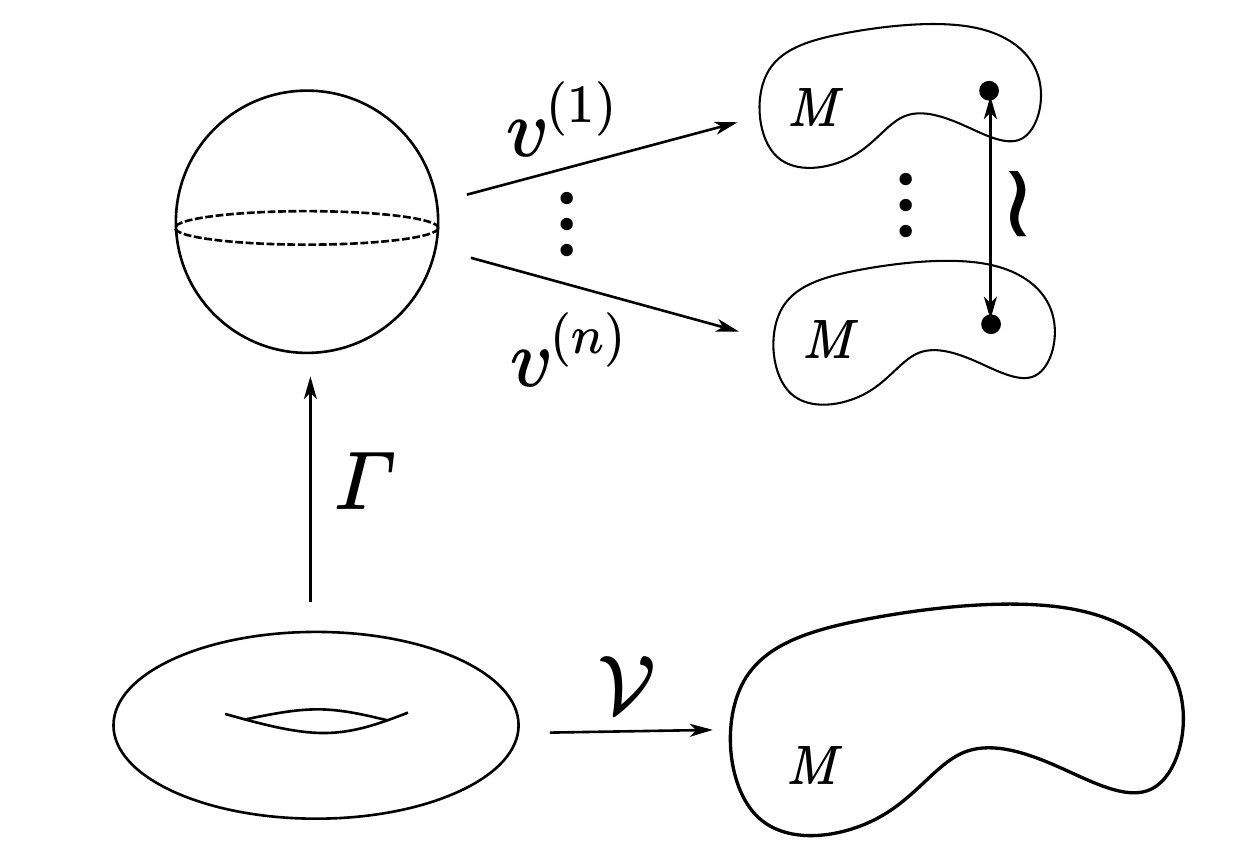}
        \caption{A sketch of the branched covering approach in orbifold correlator calculations. The fields in the twisted sector are lifted by $\Gamma$ to a single field on a covering surface $\Sigma$. Note that the genus of the covering surface can be higher than that of the original surface.} \label{fig:orbifold_covering}
\end{figure}

In a path-integral formulation of the theory, the boundary conditions on the fields can be elegantly implemented using a branched covering map. A branched covering map is a holomorphic map
\begin{equation}
    \Gamma:\ \Sigma\longrightarrow S^2\ ,
\end{equation}
where $\Sigma$ is another Riemann surface. If twist operators of cycle length $w_i$ are inserted at $x_i$, then $\Gamma$ should have ramification index $w_i$ at one point $z_i$ in the pre-image of $x_i$, i.e.\ it should be of the form 
\begin{equation}
    \Gamma(z) = x_i + a_i\,(z-z_i)^{w_i} + \mcl{O} \bigl((z-z_i)^{w_i+1}\bigr)\ ,\quad z\to z_i\ .
\end{equation}
Such a map then lifts the $v^{(k)}$ satisfying the twisted boundary conditions to a single-valued field $\mcl{V}:\,\Sigma\longrightarrow M$. More precisely, around each point $x$ that is not a twist insertion and for each $k$, there is a neighbourhood of $x$ on which $v^{(k)}$ is single valued. Then, there exists a $z_k$ with $\Gamma(z_k) = x$ and a neighbourhood of $z_k$ such that $\Gamma$ is a bi-holomorphism between the two neighbourhoods, where 
\begin{equation}\label{eq:lifted_field_relation}
    v^{(k)}(\Gamma(z)) \ \longleftrightarrow \ (\Gamma'(z))^{-h}\,\mcl{V}(z)\ ,
\end{equation}
with $h$ the conformal dimension of $v$. Here we have written ``$\leftrightarrow$'' instead of ``$=$'' because the operators technically act on different Hilbert spaces;  the arrow is meant to indicate that the application of an operator on the left in the twisted sector corresponds to the application of the one on the right in the Hilbert space of the theory on the covering surface. The situation is sketched in Figure~\ref{fig:orbifold_covering}.

The $w$-cycle twisted sector is generated from the ground state $\sigma_w(0)$ by the action of fractional modes that can be defined as follows. Let us consider the superpositions
\begin{equation}
    \mbf{v}^j(x) = \sum_{k=1}^w \omega^{j\,(k-1)}\,v^{(k)}(x)\ ,\quad j=0,\dots,w-1\ , 
\end{equation}
where $\omega = e^{2\pi i/w}$ is a primitive $w$-th root of unity. These fields satisfy the boundary conditions
\begin{equation}
    \mbf{v}^j(e^{2\pi i}\,x) = \omega^{-j}\,\mbf{v}^j(x)\ ,
\end{equation}
and thus have a mode expansion 
\begin{equation}\label{eq:general_fractional_modes}
    \mbf{v}^j(x) = 
    \sum_{r\in w\mbb{Z}+j-h\,w} v_{r/w}\,x^{-r/w-h}\ ,
\end{equation}
since all $v^{(k)}$ have conformal dimension $h$, and hence so does $\mbf{v}^j$. 
As $j$ ranges from $0$ to $w-1$, the fractional modes $v_{r/w}$ have a spacing $r\in \mbb{Z}-h\,w$. 

\subsection{Orbifold Correlators}

If $\ket{\phi}$ is a state in the $w$-twisted sector, the vertex operator associated to $v_{-r/w}\,\ket{\phi}$
equals 
\begin{equation}
V(  v_{-r/w}\,\ket{\phi}, x_0) =   \underset{C(x_0)}{\oint}\frac{dx}{2\pi i}\,\,\mbf{v}^j(x)\,(x-x_0)^{-r/w+h-1}\,V(\ket{\phi},x_0)
\end{equation}
in the orbifold, where $C(x_0)$ denotes a small contour around $x_0$ and $\mbf{v}^j$ is the field containing the mode $v_{-r/w}$ in its expansion, i.e.\  $r\in w\mbb{Z}+j-h\,w$. 
If $\Gamma(z)$ is a covering map corresponding to a correlator with the insertion of $V(  v_{-r/w}\,\ket{\phi}, x_0) $, we can change coordinates using the identification (\ref{eq:lifted_field_relation}) to lift the vertex operator to the covering surface $\Sigma$. Under this transformation, the mode $v_{-r/w}$ corresponds to an insertion of
\begin{equation}\label{eq:mode_lift}
    \underset{C(z_0)}{\oint}\frac{dz}{2\pi i}\,\,\mcl{V}(z)\,(\Gamma(z)-x_0)^{-r/w+h-1}\,(\partial \Gamma)^{1-h}
\end{equation}
on the covering surface.
For example, for $z_0 = x_0 = 0$ and $\Gamma(z) = a\,z^w$, we simply obtain 
\begin{equation}
    v_{-r/w} \ \longleftrightarrow \ w^{1-h}\,a^{-r/w}\,\mcl{V}_{-r}\ .
\end{equation}
The factor of $w$ is inconsequential, while the factor of $a$ is important, as it contains information about the other insertions inside the correlator.\footnote{Note that the covering map depends on all the fields present in the correlator.} We should stress that the mode on the covering surface $\Sigma$, eq.~(\ref{eq:mode_lift}), in general also depends on subleading terms in an expansion of $\Gamma$ around $z_0$. As we shall see in Section~\ref{sec:bps_states} below, this problem is absent for BPS states, and as a consequence the lifting of BPS states to the covering surface is particularly simple. 

\subsection{The BPS States}\label{sec:bps_states}

The symmetric orbifold theory $\text{Sym}_N(\mbb{T}^4)$ has an $\mcl{N}=4$ superconformal symmetry, and hence the BPS condition is
\begin{equation}
    h = j\ ,
\end{equation}
where $h$ is the dimension and $j$ the $\mfr{su}(2)$ spin of a state. In this section we shall describe the different BPS states of the symmetric orbifold, as well as their lift to the covering surface; our exposition will follow closely \cite{Lunin:2001pw}. 

\subsubsection{The case when \texorpdfstring{$w$}{w} is odd}

Let us first consider the case of odd $w$. Then the twisted vacuum has conformal dimension
\begin{equation}
    h_w = \frac{w^2-1}{4w}\ ,
\end{equation}
and $\mfr{su}(2)$ spin $j = 0$. The lift of the twisted sector ground state $\sigma_w(x_0)$ to the covering surface is the vacuum $\ket{0}_\mrm{NS}$, up to normalisation.

Let us denote the $\mfr{su}(2)_1$ currents of the seed theory by $J^{k,a}_x(x)$, where the subscript $x$ distinguishes these currents from their lift to the covering surface $J^a_z(z)$. One can form the fractional modes as before, see eq.~(\ref{eq:general_fractional_modes}),
\begin{equation}\label{eq:su2fractional_modes}
    J^{a}_{x, m/w} := \underset{C(0)}{\oint}\frac{dx}{2\pi i}\,\sum_{k=1}^w J^{k,a}_x(x)\,e^{2\pi\,i\,m(k-1)/w}\,x^{m/w}\ ,
\end{equation}
as the integrand is single valued, and these modes satisfy an $\mfr{su}(2)_w$ algebra
\begin{subequations}
\begin{align}
    [J^3_{x,m/w},J^3_{x,n/w}] &= -\frac{1}{2}\,w\,\frac{m}{w}\,\delta_{m+n,0}\ ,\\
    [J^{3}_{x,m/w},J^{\pm}_{x,n/w}] &= \pm J^{\pm}_{x,(m+n)/w}\ ,\\
    [J^+_{x,m/w},J^-_{x,n/w}] &= w\,\frac{m}{w}\,\delta_{m+n,0} - 2\,J^3_{x,(m+n)/w}\ .
\end{align}
\end{subequations}
The commutator with the dilation operator is
\be
    [L_0, J^{a}_{x,-m/w}] = \frac{m}{w}\,J^{a}_{x,-m/w}\ .
\ee
Hence the fractional mode $J^{+}_{x,-m/w}$ raises the $\mfr{su}(2)$ charge by $1$, while only increasing the dimension by $m/w$; thus as long as $m/w<1$, the application of $J^{+}_{x,-m/w}$ brings a state `closer to the BPS-bound'. We should mention that while the modes of $J^{a}_{x}$ form an $\mfr{su}(2)_w$ algebra, their lifts $J^a_z$ give rise to an $\mfr{su}(2)_1$ algebra on the covering surface --- this is how it should be since $\mfr{su}(2)_1$ is the symmetry of the seed theory. 

Using the fractional modes, two chiral BPS states can be found in the $w$-twisted sector, namely\footnote{Here, `chiral' means that $h=j=m$, where $m$ is the $J^3_{x,0}$ eigenvalue. The chiral BPS states are therefore $\mfr{su}(2)$ highest weight states.} 
\begin{subequations}\label{eq:BPS_states_twisted_sector}
\begin{align}
    \sigma^+_w &= J^{+}_{x,-w/w}\cdots J^{+}_{x,-3/w}J^{+}_{x,-1/w}\,\sigma_w(0)\ ,\\
    \sigma^-_w &= J^{+}_{x,-(w-2)/w}\cdots J^{+}_{x,-3/w}J^{+}_{x,-1/w}\,\sigma_w(0)\ ,
\end{align}
\end{subequations}
with $h=j=(w\pm 1)/2$, respectively. 

For the calculation in \cite{Lunin:2001pw}, as well as for the comparison with the worldsheet theory, we need the lift of these states to the covering surface. As explained in eq.~(\ref{eq:mode_lift}) above, a state $V(J^a_{x,-n/w}\,\ket{\phi},x_0)$ in the orbifold corresponds to a state
\begin{equation}\label{eq:su2_mode_lift}
    \underset{C(z_0)}{\oint}\frac{dz}{2\pi i}\,\, J^a_z(z)\, \bigl(\Gamma(z)-x_0\bigr)^{-n/w}\,V(\ket{\Phi},z_0)
\end{equation}
on the covering surface, where $V(\ket{\Phi},z_0)$ is the lift of $V(\ket{\phi},x_0)$. As a warm-up, consider first the state $J^+_{x,-1/w}\,\sigma_w(0)$ inserted at $x_0$. Near $z=z_0$ the covering map has the expansion 
\begin{equation}\label{Gammaexp}
    \Gamma(z) = x_0 + a\,(z-z_0)^w+\sum_{m=1}^\infty b_m\,(z-z_0)^{w+m}\ ,
\end{equation}
and thus
\begin{equation}\label{eq:cov_exp}
    (\Gamma(z)-x_0)^{-1/w} = \frac{a^{-\frac{1}{w}}}{z-z_0} + \sum_{m=0}^\infty c_m\,(z-z_0)^m\ .
\end{equation}
The state on the covering surface is hence
\begin{equation}
    \underset{C(z_0)}{\oint}\frac{dz}{2\pi i}\,\, (a^{-\frac{1}{w}}\,J^+_{z,-1} + \sum_{m=0}^\infty c_m\,J^+_{z,m})\,V(\ket{0}_\mrm{NS},z_0) = a^{-\frac{1}{w}}\,V(J^+_{z,-1},z_0) = a^{-\frac{1}{w}}\,J^+_z(z_0)\ ,
\end{equation}
as the $J^+_{z,m}$ modes do not contribute because they annihilate the vacuum. Similarly, if we consider the state $J^+_{x,-3/w}J^+_{x,-1/w}\,\sigma_w(0)$, the function $(\Gamma(z)-x_0)^{-3/w}$ will have a third order pole. Then the contour integral picks out $a^{-3/w}\,J^+_{z,-3}$, but in general there are also lower order contributions. For the BPS states, however, the subleading terms vanish because they are null. In particular, $J^+_{z,-1}\,J^+_{z,-1}\,\ket{0}_\mrm{NS}=0$ is null at level $k=1$, as is its $L_{-1}$ descendant, which is proportional to $J^+_{z,-2}\,J^+_{z,-1}\,\ket{0}_\mrm{NS}$.\footnote{Alternatively, this follows directly from the free field realisation in terms of free fermions.} Thus the lift of the state $J^+_{x,-3/w}J^+_{x,-1/w}\,\sigma_w(0)$ is simply 
\begin{equation}\label{step1}
    a^{-\frac{4}{w}}\,J^+_{z,-3}\,J^+_{z,-1}\,\ket{0}_\mrm{NS}\ ,
\end{equation}
inserted at $z_0$. We can iterate this argument for the chiral BPS states, and in each step only the leading pole contributes; thus  we have the simple dictionary 
\begin{equation}\label{eq:orbifold_mode_correspondence}
    J^{+}_{x,-m/w} \ \longleftrightarrow \ a^{- \frac{m}{w}}\,J^+_{z,-m}\ ,
\end{equation}
where $a$ is the leading coefficient of $\Gamma$ from eq.~(\ref{Gammaexp}).

Disregarding the normalisation of the vacuum for now, the BPS states in the $w$-cycle twisted sector (\ref{eq:BPS_states_twisted_sector}) thus correspond to 
\begin{subequations}\label{eq:BPS_states_covering_surface}
\begin{align}
    a^{-\frac{(w+1)^2}{4w}}\,J^{+}_{z,-w}\cdots J^{+}_{z,-3} J^{+}_{z,-1}\,&\ket{0}_\mrm{NS}\\
    a^{- \frac{(w-1)^2}{4w}}\,J^{+}_{z,-(w-2)}\cdots J^{+}_{z,-3} J^{+}_{z,-1}\,&\ket{0}_\mrm{NS}\ 
\end{align}
\end{subequations}
on the covering surface. We stress again that this simple lift, which only depends on the leading coefficient $a$ of $\Gamma$, is due to the special form of the BPS states; for general states also the subleading terms of $\Gamma$ in eq.~(\ref{eq:cov_exp}) will play a role.

Due to the $J^3_{x,0}$ conservation, correlators of chiral BPS states are zero. We will therefore also need to consider $\mfr{su}(2)$ descendants of these states, i.e.\ states of the form
\begin{equation}
    \big(J^{-}_{x,0}\big)^k\,\sigma^\pm_w(0)\ .
\end{equation}
Using (\ref{eq:su2_mode_lift}), one sees that on the covering surface, the application of $J^{-}_{x,0}$ corresponds to the application of $J^-_{z,0}$. Thus the corresponding states on the covering surface are $\mfr{su}(2)$ descendants of the states in eq.~(\ref{eq:BPS_states_covering_surface}). In particular, their $a$-dependence is the same.

\subsubsection{The case when \texorpdfstring{$w$}{w} is even}

For even $w$ the situation is very similar. The main difference is that the twisted sector ground states now form a doublet ($j=\frac{1}{2}$) under $\mfr{su}(2)$, and their conformal dimension equals
\begin{equation}
   h_w = \frac{w}{4}\ .
\end{equation}
Due to the parity of $w$, the states on the covering surface will be in the Ramond sector, see \cite{Lunin:2001pw}. To be specific, we shall denote by $\sigma_w$ the $m=-\frac{1}{2}$ spin down component, which is lifted to $\ket{\downarrow}_\mrm{R}$. Then the chiral BPS states in the $w$-cycle twisted sector are of the form 
\begin{subequations}
\begin{align}
    \sigma^+_w &= J^{+}_{x,-w/w}\cdots J^{+}_{x,-2/w}J^{+}_{x,0}\,\sigma_w(0)\ ,\\
    \sigma^-_w &= J^{+}_{x,-(w-2)/w}\cdots J^{+}_{x,-2/w}J^{+}_{x,0}\,\sigma_w(0)\ .
\end{align}
\end{subequations}
In the same way as above, one can find the lifts of these states to the covering surface,
\begin{subequations}\label{eq:BPS_states_covering_surface_even}
\begin{align}
    a^{-\frac{w(w+2)}{4w}}\,J^{+}_{z,-w}\cdots J^{+}_{z,-2} J^{+}_{z,0}\,&\ket{\downarrow}_\mrm{R}\ ,\\
    a^{-\frac{w(w-2)}{4w}}\,J^{+}_{z,-(w-2)}\cdots J^{+}_{z,-2} J^{+}_{z,0}\,&\ket{\downarrow}_\mrm{R}\ .
\end{align}
\end{subequations}

\section{The Worldsheet Theory}\label{sec:worldsheet}

In this section we shall review the worldsheet description of strings on $\mrm{AdS}_3\times {\rm S}^3\times \mbb{T}^4$ with $k=1$ units of NS-NS flux; this is the worldsheet theory that is exactly dual to the symmetric product orbifold of $\mbb{T}^4$. In the hybrid formalism of \cite{Berkovits:1999im} the background can be described by a WZW model based on the symmetry algebra $\mfr{psu}(1,1|2)_1$. We shall first discuss the free field realisation \cite{Dei:2020zui, Eberhardt:2019ywk} of the theory and its highest weight representations. Then we shall explain how the other representations of the worldsheet theory are obtained from the Ramond sector highest weight representation by spectral flow. Finally, we shall define the relevant worldsheet correlators and explain that they localise to those points in the moduli space where branched coverings exist \cite{Dei:2020zui}, making the duality with the orbifold manifest. Again, our aim is not to be comprehensive --- more details can be found, in particular, in \cite{Dei:2020zui, Eberhardt:2019ywk} --- but rather to fix our notation and introduce what will be needed below.

\subsection{The Free Field Algebra}

The worldsheet $\mfr{psu}(1,1|2)_1$ algebra can be realised using free fields, specifically two complex symplectic bosons denoted by $\xi^\pm$, $\eta^\pm$, and two complex fermions, $\psi^\pm$, $\chi^\pm$ with (anti)-commutation relations\footnote{All of these fields have conformal weight $h=\frac{1}{2}$.}
\begin{equation}\label{eq:free_field_CR}
[\xi^\alpha_r,\eta^\beta_s] = \epsilon^{\alpha\beta}\delta_{r+s,0}\,,\quad \{\psi^\alpha_r,\chi^\beta_s\} = \epsilon^{\alpha\beta}\delta_{r+s,0}\ ,
\end{equation}
where $\epsilon^{\alpha\beta} = -\epsilon^{\beta\alpha}$ and $\epsilon^{+-} = 1$. These fields generate a $\mfr{u}(1,1|2)_1$ algebra, see \cite{Eberhardt:2018ouy} for more details.

For the purpose of calculating the BPS correlators, the bosonic $\mfr{su}(2)_1$ subalgebra, spanned by the $K^a_m$ modes, is of some importance. In terms of the free fermions, they are given as
\begin{subequations}
\begin{align}
    K^3_m &= -\tfrac{1}{2}\,(\chi^-\psi^++\chi^+\psi^-)_m\ ,\\
    K^\pm_m &= \pm\,(\chi^\pm\psi^\pm)_m\ ,
\end{align}
\end{subequations}
with commutation relations
\begin{subequations}
\begin{align}\label{eq:su2_1}
[K^3_m,K^\pm_n] &= \pm K^\pm_{m+n}\ ,\\
[K^3_m, K^3_n] &= \tfrac{1}{2}\,m\,\delta_{m+n,0}\ ,\\
[K^+_m, K^-_n] &= m\,\delta_{m+n,0} + 2\,K^3_{m+n}\ .
\end{align}
\end{subequations}

\subsection{The Highest Weight Representations}\label{sec:vacuum_representation}

The $\mfr{u}(1,1|2)_1$ algebra has only two highest weight representations: the NS vacuum representation, in which all free fields are half-integer moded, and that is generated from a single state $\ket{0}$, satisfying 
\be
\xi^\pm_r\, \ket{0} = \eta^\pm_r\, \ket{0} = \psi^\pm_r\, \ket{0} = \chi^\pm_r\, \ket{0} = 0 \ , \qquad r>0 \ ,
\ee
by the action of the negative modes. The other representation is the Ramond sector representation, in which all modes are integer moded. On the ground states the zero modes of the symplectic bosons may be taken to act as \cite{Dei:2020zui}
\begin{subequations}
\begin{align}
\xi_0^+\ket{m_1,m_2} &= \ket{m_1,m_2+\tfrac{1}{2}}\ ,& \eta_0^+\ket{m_1,m_2} &= 2\,m_1\,\ket{m_1+\tfrac{1}{2},m_2}\ ,\\
\xi_0^-\ket{m_1,m_2} &= -\ket{m_1-\tfrac{1}{2},m_2}\ ,& \eta_0^-\ket{m_1,m_2} &= -2\,m_2\,\ket{m_1,m_2-\tfrac{1}{2}}\ .
\end{align}
\end{subequations}
Then we have 
\begin{equation}
J_0^3\ket{m_1,m_2} = (m_1+m_2)\,\ket{m_1,m_2}\ ,
\end{equation}
while the Casimir of $\mathfrak{sl}(2,\mbb{R})$ is given by $C^{\mfr{sl}(2,\mbb{R})} = -j(j-1)$ with 
$j = m_1-m_2$.

Including the fermions and choosing $\chi_0^+\ket{m_1,m_2} = \psi_0^+\ket{m_1,m_2} = 0$, one finds for each $\ket{m_1,m_2}$ a $\mfr{su}(2)$ doublet spanned by
\begin{equation}
\ket{m_1,m_2}\ ,\quad \chi_0^-\psi_0^-\ket{m_1,m_2}\ ,
\end{equation}
and two singlets, spanned by
\begin{equation}
\chi_0^-\ket{m_1,m_2}\ ,\quad \psi_0^-\ket{m_1,m_2}\ ,
\end{equation}
respectively. The $Z_0=0$ condition, reducing $\mfr{u}(1,1|2)_1$ to $\mfr{psu}(1,1|2)_1$, see \cite{Dei:2020zui}, then forces $j=1/2$ for the doublet and $j=0$ (or $j=1$) for the singlets. Explicitly, the doublets are described by
\begin{equation}\label{notation}
\ket{m_1,m_2}\otimes \ket{\uparrow}_\mrm{R}\ ,\quad \ket{m_1,m_2}\otimes \ket{\downarrow}_\mrm{R}\ ,
\end{equation}
with $m_1-m_2 = 1/2$. These doublets will play an important role below.

\subsection{Spectral Flow}

An essential ingredient of the worldsheet theory is that it does not only contain highest weight representations, but also representations that are obtained from them by spectral flow \cite{Maldacena:2000hw}, see also \cite{Henningson:1991jc}. These spectrally flowed representations are in general not highest weight. In terms of the free fields there are two kinds of spectral flow,
\begin{subequations}\label{eq:spectral_flow_free_fields}
\begin{align}
\sigma^{(+)}(\eta^+_r) &= \eta^+_{r-1/2}\ ,& \sigma^{(-)}(\xi^+_r) &= \xi^+_{r-1/2}\ ,\\
\sigma^{(+)}(\xi^-_r) &= \xi^-_{r+1/2}\ ,& \sigma^{(-)}(\eta^-_r) &= \eta^-_{r+1/2}\ ,\\
\sigma^{(+)}(\chi^+_r) &= \chi^+_{r+1/2}\ ,& \sigma^{(-)}(\psi^+_r) &= \psi^+_{r+1/2}\ ,\\
\sigma^{(+)}(\psi^-_r) &= \psi^-_{r-1/2}\ ,& \sigma^{(-)}(\chi^-_r) &= \chi^-_{r-1/2}\ ,
\end{align}
\end{subequations}
and they induce a spectral flow for  $\mfr{u}(1,1|2)_1$ via $\sigma = \sigma^{(+)}\circ\ \sigma^{(-)}$. 
On the bosonic subalgebra this combined flow takes the form 
\begin{subequations}\label{eq:spectral_flow}
\begin{align}
\sigma^w(J^3_m) &= J^3_m + \tfrac{1}{2}\,w\,\delta_{m,0}\ ,\\
\sigma^w(J^\pm_m) &= J^\pm_{m\mp w}\ ,\\
\sigma^w(K^3_m) &= K^3_m + \tfrac{1}{2}\,w\,\delta_{m,0}\ ,\\
\sigma^w(K^\pm_m) &= K^\pm_{m\pm w}\ ,
\end{align}
\end{subequations}
while for the energy-momentum tensor we find 
\begin{equation}
\sigma^w(L_m) = L_m + w\,(K^3_m - J^3_m)\ .
\end{equation}
Spectral flow defines an automorphism of $\mfr{u}(1,1|2)_1$ (and similarly for $\mfr{psu}(1,1|2)_1$), and it induces an action on any representation. More specifically, we take the $w$-spectrally flowed representation to be spanned by the states of the form $[\Phi]^{\sigma^w}$, where $\Phi$ is an arbitrary state of the Ramond sector representation, and the action of a generator of $\mfr{u}(1,1|2)_1$ is defined via 
\begin{equation}
A\,[\Phi]^{\sigma^w} := [\sigma^w(A)\,\Phi]^{\sigma^w}\ .
\end{equation}

\subsection{Physical States}\label{sec:hybrid_states}

In the hybrid formalism of \cite{Berkovits:1999im}, the $\mfr{psu}(1,1|2)_1$ WZW model describes the ${\rm AdS}_3 \times {\rm S}^3$ part of the background. The worldsheet theory contains, in addition, a topologically twisted $\mathbb{T}^4$, as well as various ghost fields. The physical states can be characterised as being ${\cal N}=4$ topological; more specifically, this means that they satisfy\footnote{In the free field realisation this physical state condition was recently studied in some detail in \cite{Gaberdiel:2022bfk}.} 
\begin{equation}\label{eq:physical_state_conditions}
    G_0^+\,\psi = \widetilde{G}_0^+\,\psi = (J_0-1)\,\psi = T_0\,\psi = 0\ ,
\end{equation}
where $G^+$, $\widetilde{G}^+$, $J$ and $T$ are (some of) the ${\cal N}=4$ generators which can be constructed out of the above fields, for more detail see e.g.\ \cite{Berkovits:1999im,Gaberdiel:2022bfk}. (In addition, physical states that differ by BRST-exact states are to be identified.) A useful ansatz for the physical states \cite{Gerigk:2012cq, Gaberdiel:2021njm, Dei:2020zui} is
\begin{equation}\label{eq:physical_state_ansatz}
    \psi = \Phi\,e^{2\,\rho + i\,\sigma + i\,H}\ ,
\end{equation}
where $\Phi$ is a state in the $\mfr{psu}(1,1|2)_1$ theory. The fields $\rho\,,\sigma$ bosonise the ghosts, and $H=H_1+H_2$ is related to the (twisted) $\mbb{T}^4$ fermions by bosonisation. The condition that $\psi$ in (\ref{eq:physical_state_ansatz}) is in fact physical requires that $\Phi$ satisfies
\begin{equation}\label{eq:reduced_physical_state_conditions}
    \mcl{Q}_n\,\Phi = L_n\,\Phi = 0\ ,\quad \forall\,\,n\geq 0\ ,
\end{equation}
where $L$ is the $\mfr{psu}(1,1|2)_1$ stress-energy tensor and
\begin{equation}
\mcl{Q} = 2\,(\chi^+\chi^-)\,(\xi^+\partial\xi^--\xi^-\partial\xi^+)\ .
\end{equation}

\subsection{Correlators in the Hybrid Formalism}\label{sec:hybrid_subtleties}

As a topological ${\cal N}=4$ string theory, the definition of the correlators involves additional insertions of supercharges  \cite{Berkovits:1999im}. On the physical states of the form (\ref{eq:physical_state_ansatz}), this amounts to inserting the operator $\mcl{Q}$ \cite{Dei:2020zui}. Alternatively, we may apply $\mcl{Q}_{-1}$ directly to $n-2+2\,g$ states in the correlator, where $g$ is the genus of the worldsheet. If one considers vertex operators of spectrally flowed highest weight states $\ket{m_1,m_2}$, $\mcl{Q}_{-1}$ acts as $m_1\mapsto m_1 -1/2$, $m_2\mapsto m_2 + 1/2$, see \cite{Dei:2020zui}.

An additional complication arises in the free field realisation of $\mfr{u}(1,1|2)_1$ since $\mcl{Q}$ carries non-trivial $U_0$ charge. In order to compensate for this, it was proposed in \cite{Dei:2020zui} that $n-2+2\,g$
`vacuum fields' 
\begin{equation}
W(z) := V(\ket{0}^{(1)}, z)\ ,
\end{equation}
need to be inserted, where
\begin{equation}
\ket{0}^{(1)} = [\psi^+_{-3/2}\psi^-_{-3/2}\psi^+_{-1/2}\psi^-_{-1/2}\ket{0}]^{(\sigma^{(+)}\circ(\sigma^{(-)})^{-1})^2}
\end{equation}
is the vacuum with respect to the $\mfr{psu}(1,1|2)_1$ algebra (but carries non-trivial $U_0$ charge).

\subsection{Bosonic Correlators and the Incidence Relation}

The $\mfr{sl}(2,\mbb{R})$ subalgebra of $\mfr{psu}(1,1|2)$ can be identified with M\"obius generators of the spacetime CFT, and this allows us to introduce a spacetime dependence $x$ for the vertex operators associated to physical states via \cite{Kutasov:1999xu,Maldacena:2001km,Eberhardt:2019ywk}
\be
V(\psi;x,z) = e^{x\,J^+_0}\, V(\psi;x=0,z) \, \,e^{-x\,J^+_0} \ .
\ee
In \cite{Eberhardt:2019ywk,Dei:2020zui} only the vertex operators corresponding to the twisted sector ground states were considered, and they were denoted by $V^w_{m_1,m_2}(x;z)$, where $w$ denotes the spectral flow, while $m_1$ and $m_2$ label the Ramond sector ground states before spectral flow \cite{Dei:2020zui}
\begin{equation}
    V^w_{m_1,m_2}(x;z) := V([\ket{m_1,m_2}]^{\sigma^w}, x; z)\ .
\end{equation}
It was shown in \cite{Dei:2020zui} that their correlators on the worldsheet sphere ($g=0$)
\begin{equation}\label{eq:bosonic_corr}
    \left\langle \prod_{\alpha = 1}^{n-2} W(u_\alpha)\, \prod_{i=1}^n V^{w_i}_{m_1^i,m_2^i}(x_i;z_i) \right\rangle\ ,
\end{equation}
are localised to those points in moduli space where a holomorphic covering map $\Gamma(z)$ exists. Here, $\Gamma:\,S^2\longrightarrow S^2$ is characterised by its degree and the property that 
\begin{equation}\label{eq:covering_ramification}
    \Gamma(z) = x_i + a_i^\Gamma\,(z-z_i)^{w_i} + \mcl{O} \bigl((z-z_i)^{w_i+1}\bigr)\,,\quad z\to z_i\ ,
\end{equation}
for all $i=1,\dots, n$. For any choice of $w_i$ and $x_i$, there are only finitely many choices of $z_i$ for which such a map exists --- in particular, this implies that the worldsheet correlator localises. This localisation property could be derived from the identity (which in turn was deduced from the OPEs of $\xi^\pm$ with the $V^w_{m_1,m_2}$) \cite{Dei:2020zui}
\begin{equation}\label{eq:incidence_relation}
    \left\langle \Bigl(\xi^-(z)+\Gamma(z)\,\xi^+(z)\Bigr)\, \prod_{\alpha = 1}^{n-2} W(u_\alpha)\, \prod_{i=1}^n V^{w_i}_{m_1^i,m_2^i}(x_i;z_i) \right\rangle = 0\ ,
\end{equation}
provided that a branched covering exists. If no branched covering exists, the correlator is zero. As a consequence, the correlators take the form 
\begin{equation}\label{eq:symplectic_boson_correlator}
    \left\langle \prod_{\alpha = 1}^{n-2} W(u_\alpha)\, \prod_{i=1}^n V^{w_i}_{m_1^i,m_2^i}(x_i;z_i) \right\rangle\ = \sum_\Gamma W_\Gamma(z_i,u_\alpha,j_i)\,\prod_{i=1}^n\,(a_i^\Gamma)^{-h_i}\,\prod_{i = 4}^n \delta(x_i-\Gamma(z_i))\ ,
\end{equation}
where $j_i = m_1^i-m_2^i$ is the $\mrm{sl}(2,\mbb{R})$ spin, $h_i = m_1^i+m_2^i+w_i/2$ is the worldsheet conformal dimension, and $a_i^\Gamma$ denotes the coefficients in (\ref{eq:covering_ramification}).

While the incidence relation (\ref{eq:incidence_relation}) allows one to deduce this localisation property, it does not quite fix the complete $h_i$ dependence of (\ref{eq:symplectic_boson_correlator}). Finding the full exponent with which $a_i$ appears is much more difficult, and only the case of three- and four-point correlators has so far been worked out in detail \cite{Dei:2021xgh,Dei:2021yom,Dei:2022pkr}; in this case, the $a_i$ dependence of the correlator was found to equal
\begin{equation}
    a_i^{-h_i + (w_i-1)/4}\ .
\end{equation}
Furthermore, there are additional factors depending on the residues of the poles of $\Gamma$ that guarantee that the $x_i$ dependence of the correlator is correct.\footnote{For example, for a $3$-point function this is fixed by the conformal symmetry.} We will ignore these pole coefficients and only consider the structure in the $a_i$. We should also mention that while we have so far concentrated on the case where the worldsheet is a sphere, the generalisation to higher genera is also possible, see \cite{Eberhardt:2020akk,Knighton:2020kuh}.

\subsection{The Worldsheet BPS States}\label{sec:dual_bps_states}

We close this section by identifying the worldsheet states that are dual to the BPS states of Section~\ref{sec:bps_states} in the symmetric orbifold. 

In the worldsheet description, the $\mfr{sl}(2,\mbb{R})$ algebra generated by $J^a_0$ corresponds to the M\"obius algebra on the boundary. Furthermore, the $\mfr{su}(2)$ algebra generated by $K^a_0$ corresponds to the $\mfr{su}(2)$ subalgebra of the $\mcl{N} = 4$ superalgebra in the dual CFT, as follows from the construction of DDF operators in \cite{Giveon:1998ns,Eberhardt:2019qcl,Naderi}, see also \cite{Gaberdiel:2021njm} for more explicit checks. We are therefore looking for physical states in the $w$ spectrally flowed sector with $K^3_0$ and $J^3_0$ eigenvalue $(w\pm 1)/2$. Since under spectral flow both the $J^3_0$ and the $K^3_0$ eigenvalues are shifted by $\frac{w}{2}$, see eq.~(\ref{eq:spectral_flow}), two candidate states are
\begin{equation}\label{eq:hw_BPS_states}
[\Phi^+]^{\sigma^w} = \left[\ket{1/2,0}\otimes \ket{\uparrow}\right]^{\sigma^w}\,,\qquad [\Phi^-]^{\sigma^w} = \left[\ket{0, -1/2}\otimes \ket{\downarrow}\right]^{\sigma^w}\ ,
\end{equation}
where we have used the notation from eq.~(\ref{notation}).  It is not difficult to check that these states also satisfy the physical state conditions of eq.~(\ref{eq:reduced_physical_state_conditions}), and hence give rise to physical states when combined with the ghost factors as in (\ref{eq:physical_state_ansatz})
\begin{equation}
    [\Phi^\pm]^{\sigma^w}\,e^{2\,\rho+i\,\sigma +i\,H} \ . 
\end{equation}
Obviously, the same is true for any $K_0^-$ descendant. Thus, the bosonic part of the three-point correlators (\ref{eq:bosonic_corr}) will be exactly of the form studied in \cite{Dei:2020zui}, satisfying the incidence relation (\ref{eq:incidence_relation}).
For the following it will be convenient to undo the spectral flow for the fermions. Let us first concentrate on the case where $w$ is odd. By the action of the fermionic modes on the doublet $\ket{\uparrow},\ket{\downarrow}$, it follows from the spectral flow operation of eq.~(\ref{eq:spectral_flow_free_fields}) that 
\begin{equation}
\psi^+_{-r}\,[\Phi^+]^{\sigma^w} = \chi^+_{-r}\,[\Phi^+]^{\sigma^w} = 0\ ,\quad r \in \{1/2,\dots,w/2\}\ ,
\end{equation}
and similarly
\begin{equation}
\psi^+_{-s}\,[\Phi^-]^{\sigma^w} = \chi^+_{-s}\,[\Phi^-]^{\sigma^w} = 0\ ,\quad s \in \{1/2,\dots,(w-2)/2\}\ .
\end{equation}
This characterises the fermionic part of the state completely, and we can rewrite eq.~(\ref{eq:hw_BPS_states}) as 
\begin{subequations}\label{eq:dual_bps_states_full}
\begin{align}
[\Phi^+]^{\sigma^w} &= [\ket{1/2,0}]^{\sigma^w}\otimes \bigl( K^+_{-w}\cdots K^+_{-1}\,\ket{0}_\mrm{NS} \bigr)\ ,\\
[\Phi^-]^{\sigma^w} &= [\ket{0,-1/2}]^{\sigma^w}\otimes \bigl(K^+_{-(w-2)}\cdots K^+_{-1}\,\ket{0}_\mrm{NS}\bigr)\ ,
\end{align}
\end{subequations}
where the spectral flow now only acts on the $\mfr{sl}(2,\mbb{R})$ part of the theory.
Comparing this to the orbifold state on the covering space, see eq.~(\ref{eq:BPS_states_covering_surface}), we observe that exactly the same $\mfr{su}(2)_1$ descendants appear. However, in the lift of the orbifold, we also had some factors of the covering map coefficient $a_i$, which the free fermions on the worldsheet do not see. As we will discuss below, these factors will be accounted for by the bosonic part of the worldsheet correlator.

For even $w$, the analysis is similar, and the actual BPS states take the form 
\begin{subequations}\label{eq:dual_bps_states_full_even}
\begin{align}
[\Phi^+]^{\sigma^w} &= [\ket{1/2,0}]^{\sigma^w}\otimes \bigl( K^+_{-w}\cdots K^+_{-2}\,K^+_{0}\,\ket{\downarrow}_\mrm{R} \bigr)\ ,\\
 [\Phi^-]^{\sigma^w} &= [\ket{0,-1/2}]^{\sigma^w}\otimes \bigl( K^+_{-(w-2)}\cdots K^+_{-2}\,K^+_{0}\,\ket{\downarrow}_\mrm{R}\bigr) \ .
\end{align}
\end{subequations}
Again, this has the same $\mfr{su}(2)_1$ descendant structure as the lifts to the covering surface of the BPS states in the symmetric orbifold, see eq.~(\ref{eq:BPS_states_covering_surface_even}). 

We should stress that in either case the $\mfr{su}(2)_1$ descendants of the lifted orbifold states match those on the worldsheet. This provides a concrete realisation of the idea that the worldsheet states can be identified with the lift of the orbifold states --- this must in a sense be true if the worldsheet plays indeed the role of the covering surface,  as originally suggested in \cite{Lunin:2000yv,Lunin:2001pw,Pakman:2009zz,Pakman:2009ab} and later partially confirmed in \cite{Eberhardt:2019ywk, Dei:2020zui}.

\section{The BPS Correlators}\label{sec:bps_correlators}

In this section we shall explain that the correlators of the BPS states in the symmetric orbifold theory agree with those calculated from a worldsheet perspective. The key observation is that the lift of a BPS state of the symmetric orbifold to the covering surface agrees, as regards its $\mfr{su}(2)_1$ descendant structure, with the corresponding worldsheet dual state, compare eqs.~(\ref{eq:BPS_states_covering_surface}) and (\ref{eq:dual_bps_states_full}), or eqs.~(\ref{eq:BPS_states_covering_surface_even}) and (\ref{eq:dual_bps_states_full_even}). As we shall explain, this is sufficient to deduce that their correlators must agree up to an overall factor.\footnote{This factor is the contribution which is not fixed by the incidence relation (\ref{eq:incidence_relation}), and is thus dependent on the cycle length of the twist operators and their insertion points.} We shall also exemplify this finding by an explicit calculation, see Appendix~\ref{sec:app_calculations}. 

\subsection{The symmetric orbifold calculation}

Let us first review the calculation of the BPS correlators for the symmetric orbifold. For a given choice of $(x_i,w_i)$, there exist only finitely many branched covering maps, i.e.\ holomorphic maps that behave as \begin{equation}\label{eq:covering}
    \Gamma(z) = x_i + a_i\,(z-z_i)^{w_i} + \mcl{O}\bigl((z-z_i)^{w_i+1}\bigr)\ ,\quad z\to z_i\ 
\end{equation}
near $z=z_i$, and are only ramified at these points, i.e.\ $\partial\Gamma(z)=0$ only at $z=z_i$. As we explained in Section~\ref{sec:bps_states}, the lift of a BPS state to the covering surface only depends on the leading behaviour of a branched covering $\Gamma$ near the ramification points, i.e.\ on  the coefficient $a_i$ in (\ref{eq:covering}). 
The BPS correlator on the covering surface $\Sigma$ then involves $\mfr{su}(2)_1$ descendants of the highest weight states in eqs.~(\ref{eq:BPS_states_covering_surface}) and (\ref{eq:BPS_states_covering_surface_even}). Keeping track of the various factors of $a_i$, and remembering that there is a contribution 
\be
a_i^{-(w^2_i-1)/(4w_i) + \frac{w_i-1}{4}}
\ee
due to the conformal anomaly \cite{Lunin:2000yv}, the total exponent of $a_i$ becomes, for $w_i$ odd, 
\be\label{eq:cov_map_coeff_exponent}
- \frac{(w_i\pm 1)^2}{4w_i} - \frac{w^2_i-1}{4w_i} + \frac{w_i-1}{4} = - \frac{w_i\pm 1}{2} + \frac{w_i-1}{4} \ ,
\ee 
where the sign depends on which of the two BPS states in the $w$-cycle twisted sector is being considered, see eq.~(\ref{eq:BPS_states_covering_surface}). Finally, we need to sum over all the possible branched covering maps. Note that there can be contributions from disconnected covering surfaces if the twist fields permute independent copies \cite{Dei:2019iym}. In this case, the correlator factors into the connected correlators, and it is thus sufficient to study the connected contributions.

For even $w_i$, there is an additional contribution due to the insertion of $\ket{\downarrow}_\mrm{R}$. The Ramond vacua are created from the Neveu-Schwarz vacuum by the spin operators $\mcl{S}^\updownarrow$, which have conformal dimension $1/4 = c/24$. Therefore, an insertion of $\mcl{S}^\downarrow$ comes with a factor of $a_i^{-1/(4\,w_i)}$ on the covering surface, and the analogue of eq.~(\ref{eq:cov_map_coeff_exponent}) for $w_i$ even is 
\begin{equation}
    - \frac{w_i(w_i\pm 2)}{4w_i} - \frac{1}{4\,w_i} - \frac{w^2_i-1}{4w_i} + \frac{w_i-1}{4} =-\frac{w_i\pm 1}{2} + \frac{w_i-1}{4}\ .
\end{equation}
This then also agrees with eq.~(\ref{eq:cov_map_coeff_exponent}).

\subsection{The worldsheet calculation}\label{sec:worldsheet_calculation}

We can now compare these results to the worldsheet correlator. In the hybrid formalism, once the correct supercharges are applied, see Section \ref{sec:hybrid_subtleties}, we need to integrate over the moduli spaces $\mcl{M}_{g,n}$ of punctured Riemann surfaces. As the dual BPS states (\ref{eq:dual_bps_states_full}) have a symplectic boson part of the form studied in \cite{Dei:2020zui}, the incidence relation (\ref{eq:incidence_relation}) shows that the worldsheet correlator localises to those points in moduli space where a branched covering map $\Gamma$ exists. This holds for all genera of the worldsheet \cite{Eberhardt:2020akk,Knighton:2020kuh}. Furthermore, from (\ref{eq:symplectic_boson_correlator}), it follows that a factor of
\begin{equation}
    a_i^{-(w_i\pm 1)/2}
\end{equation}
appears in the worldsheet correlator. We are thus left with the same $\mfr{su}(2)_1$ correlator, evaluated over the same covering surface,\footnote{As mentioned in the previous section, in the orbifold there can be disconnected covering surfaces. Our worldsheet calculation only gives the contribution of the connected surfaces, but it is clear that the disconnected contributions are correctly reproduced if we go to a second quantised formulation.} and hence we manifestly reproduce the symmetric orbifold calculation, up to the symplectic boson prefactor not fixed by the incidence relation. If we assume that the ground state correlators agree with the orbifold result, as has been shown for three- and four-point correlators \cite{Dei:2021xgh,Dei:2021yom,Dei:2022pkr}, then the identical $\mfr{su}(2)_1$ structure implies that the BPS correlators also agree. 

There is one small additional subtlety that needs to be addressed: the worldsheet correlator also contains additional 
 insertions of the $W$ and $\mcl{Q}$ fields, see Section \ref{sec:hybrid_subtleties}, and one may be worried that they modify the result. However, with respect to the $\mfr{su}(2)_1$ algebra, both fields behave as the vacuum: for $W$ this is the case by construction since it behaves as the vacuum with respect to $\mfr{psu}(1,1|2)_1$, and for the $\mcl{Q}$ field, this is a consequence of the $\chi^+\chi^-$ term that defines a $\mfr{su}(2)_1$  singlet. The $W$ and $\mcl{Q}$ fields thus can only contract with each other and they just give rise to an overall prefactor. Note that these arguments hold for arbitrary genus, since they only use localisation, the behaviour of the leading order terms of the covering map, and the $\mfr{su}(2)_1$ structure of the dual BPS states. Therefore, the \textit{full} correlators (i.e.\ including the sum over all covering maps) match between the two sides.

We can also test these ideas more explicitly. In \cite{Lunin:2001pw}, many three-point functions of BPS operators in the symmetric product orbifold were calculated (for genus zero), and a general result was conjectured. As we explain in Appendix~\ref{sec:app_calculations}, we have managed to reproduce ratios of their conjectured result (that are independent of the overall normalisation ambiguity)  from a worldsheet calculation.

\subsection{Formal Generalisation}\label{sec:formal_ddf_operators}

The above argument hinges on the fact that the worldsheet state corresponding to a given BPS state $\phi$ in the symmetric orbifold theory can be identified with the lift of $\phi$ to the covering surface. This then implies that the worldsheet correlator is essentially the same as the correlator calculated on the covering surface. Given the generality of the argument, one may suspect that something similar should hold in general, not just for the BPS states. 

The situation is, however, a little bit more complicated. The covering map depends, in general, on all the fields in the correlator, and hence so does the lift to the covering surface. For the BPS states the lift to the covering surface only depends on the $a_i$ parameters, see the discussion in Section~\ref{sec:bps_states}, and this factor appeared, from the worldsheet perspective, from the correlators of the symplectic boson states, see in particular eq.~(\ref{eq:symplectic_boson_correlator}).  For a general state, however, also the subleading terms in eq.~(\ref{eq:cov_exp}) will contribute to the lift to the covering surface, and it is then not obvious how they will be reproduced from the worldsheet perspective. 

In the following we shall argue that this difficulty is resolved once we consider the correct physical states of the worldsheet theory. More specifically, the worldsheet dual of an arbitrary descendant state of the symmetric orbifold can be described in terms of the so-called DDF operators on the worldsheet. These operators depend on a field $\gamma$, which inside any correlator behaves as the covering map; as a consequence the DDF operators reproduce precisely the lifting of $\phi$ to the covering surface.

\subsubsection{The Wakimoto Representation}

The DDF operators \cite{DelGiudice:1971yjh,Giveon:1998ns,Eberhardt:2019qcl} are most easily constructed using the Wakimoto representation \cite{Wakimoto:1986gf}. For $\mfr{sl}(2,\mbb{R})_1\subseteq \mfr{psu}(1,1|2)_1$ the Wakimoto representation takes the form 
\begin{subequations}\label{eq:Wakimoto}
\begin{align}
    J^+ &= \beta\ ,\\
    J^3 &= (\beta\gamma) - \partial \Phi\ ,\\
    J^- &= (\beta\gamma\gamma) -2\,\partial\gamma - 2\,(\partial\Phi\,\gamma)\ .
\end{align}
\end{subequations}
Here, $\Phi$ is a timelike free boson with background charge $1$, and $\beta$ and $\gamma$ are fields of dimension $1$ and $0$, respectively, with OPE
\begin{equation}
    \beta(z)\,\gamma(w) \sim -\frac{1}{z-w}\ .
\end{equation}
The field $\gamma$ acts on the spectrally flowed ground states as
\begin{equation}
    \gamma(z)\,[\ket{m_1,m_2}]^{\sigma^w} = z^w\,[\ket{m_1-1/2,m_2-1/2}]^{\sigma^w} + \mcl{O}(z^{w+1})\ ,\quad z\to 0\ .
\end{equation}
Thus, $\gamma^{1/w}$ is a well-defined field in the $w$-twisted sector \cite{Eberhardt:2019qcl}. It furthermore satisfies 
\begin{equation}\label{eq:gammax}
    e^{x\,J_0^+}\,\gamma(z)\,e^{-x\,J_0^+} = \gamma(z) - x\ ,
\end{equation}
as follows directly from the above free field realisation, see eq.~(\ref{eq:Wakimoto}). Using the incidence relation of eq.~(\ref{eq:incidence_relation}) together with the $m_1^i,m_2^i$ dependence of the correlator of eq.~(\ref{eq:symplectic_boson_correlator}),  it follows that 
\begin{equation}\label{eq:leading_order_agreement}
    \frac{\Big\langle \gamma(z)\,\prod_{\alpha=1}^{n-2}W(u_\alpha)\,\prod_{i=1}^n V^{w_i}_{m_1^i,m_2^i}(x_i;z_i) \Big\rangle}{\Big\langle \prod_{\alpha=1}^{n-2}W(u_\alpha)\,\prod_{i=1}^n V^{w_i}_{m_1^i,m_2^i}(x_i;z_i) \Big\rangle} = x_i + a_i\,(z-z_i)^{w_i} + \mcl{O}((z-z_i)^{w_i+1})\ ,\,\, z\to z_i\ ,
\end{equation}
i.e., inside a correlator, $\gamma$ agrees with the covering map to leading order. This suggests the identity
\begin{equation}\label{eq:gammacov}
    \Big\langle \gamma(z)\,\prod_{\alpha=1}^{n-2}W(u_\alpha)\,\prod_{i=1}^n V^{w_i}_{m_1^i,m_2^i}(x_i;z_i) \Big\rangle = \Gamma(z)\,\Big\langle \prod_{\alpha=1}^{n-2}W(u_\alpha)\,\prod_{i=1}^n V^{w_i}_{m_1^i,m_2^i}(x_i;z_i) \Big\rangle \ , 
\end{equation}
which was in fact confirmed experimentally also to subleading order \cite{Eberhardt:2019ywk}. We should mention that while (\ref{eq:gammacov}) appears very natural, it predicts that the correlator on the left has poles in $z$ at positions where there are no operator insertions --- this simply follows from the fact that the covering map $\Gamma(z)$ has such poles. Thus the correlator must have additional field insertions that give rise to these poles, but are invisible from the perspective of $\mathfrak{psu}(1,1|2)_1$. While it was shown in  \cite[Section~6.2]{Eberhardt:2019ywk} that such operators do indeed exist, the question of why these operators are present and inserted at the correct positions has not yet been understood satisfactorily. This issue may also be related to the question of whether screening operators are required, see \cite{Giribet:2000fy,Hosomichi:2000bm}.

\subsubsection{The DDF Operators}

Using the $\gamma$ field, we can now construct the DDF operators via
\begin{equation}\label{eq:ddf_operators}
    \mcl{K}^a_{r} = \underset{C(0)}{\oint} \frac{dz}{2\pi i}\,\,K^a(z)\,\gamma^r(z)\ ,
\end{equation}
which formally looks like a coordinate transformation $z\mapsto \gamma(z)$. In order to make sense of these operators we need to define the operator $\gamma^r$ for negative $r$; one way to do so was recently introduced by Naderi \cite{Naderi} in terms of bosonised symplectic bosons. In particular, he showed that the operators (\ref{eq:ddf_operators}) indeed commute with the physical state conditions (\ref{eq:physical_state_conditions}), and that they are therefore DDF operators.\footnote{Note that these operators are seemingly simpler than those found in \cite{Giveon:1998ns,Eberhardt:2019qcl}. The reason for this is that eq.~(\ref{eq:ddf_operators}) is defined in the hybrid formalism, while the DDF operators of \cite{Giveon:1998ns,Eberhardt:2019qcl} were formulated in the RNS formalism.}

In the $w$-twisted sector, the moding is allowed to be fractional with integer multiples of $\frac{1}{w}$. Moreover, the $\mcl{K}^a_r$ modes form the algebra $\mfr{su}(2)_w$ \cite{Giveon:1998ns}, and hence can be naturally identified with the twisted orbifold algebra \cite{Eberhardt:2019qcl}. Finally, it follows from eq.~(\ref{eq:gammax}) that 
\begin{equation}
    [J_0^+,\gamma^r] = -r\,\gamma^{r-1} = -\frac{\partial}{\partial \gamma}\,\gamma^r\ ,
\end{equation}
see also \cite{Roumpedakis:2018tdb}. Thus we have 
\begin{equation}
    e^{x\,J_0^+}\,\gamma^r(z)\,e^{-x\,J_0^+} = (\gamma(z) - x)^r\ ,
\end{equation}
and we can rewrite $V(\mcl{K}^a_{-n/w}\,\ket{\phi}, x_0; z_0)$ as
\begin{align}
          V(\mcl{K}^a_{-n/w}\,\ket{\phi}, x_0; z_0) &= e^{x_0\,J_0^+}\,e^{z_0\,T_{-1}}\!\!\underset{C(0)}{\oint}\frac{dz}{2\pi i}\,\,K^a(z)\,\gamma^{-n/w}(z)\,V(\ket{\phi},0;0)\,e^{-z_0\,T_{-1}}\,e^{-x_0\,J_0^+} \nonumber \\
        &= \underset{C(z_0)}{\oint}\frac{dz}{2\pi i}\,\,e^{x_0\,J_0^+}\,K^a(z)\,\gamma^{-n/w}(z)\,e^{-x_0\,J_0^+}\,V(\ket{\phi},x_0;z_0) \nonumber \\
        &= \underset{C(z_0)}{\oint}\frac{dz}{2\pi i}\,\,K^a(z)\,(\gamma(z) - x_0)^{-n/w}\,V(\ket{\phi},x_0;z_0)\ .  
 \end{align}
This is the complete analogue of the lift of the twisted modes in the symmetric orbifold to the covering surface, see eq.~(\ref{eq:su2_mode_lift}). Thus if we assume that $\gamma^r = \Gamma^r$ in the sense of eq.~(\ref{eq:gammacov}), our worldsheet correlators reproduce exactly the symmetric orbifold correlators as evaluated on the covering surface.

\subsubsection{A consistency check}

As a small consistency check of this proposal let us confirm that the above prescription for the dual worldsheet states agrees with the result for the BPS states found in Section~\ref{sec:dual_bps_states}. We begin by noting that the leading term in the action of $\gamma^r$ on a spectrally flowed ground state is
\begin{equation}
    \gamma^r(z)\,[\ket{m_1,m_2}]^{\sigma^w} = z^{r\,w}\,[\ket{m_1-\tfrac{r}{2},m_2-\tfrac{r}{2}}]^{\sigma^w} + \mcl{O}(z^{r\,w+1})\ ,\quad z\to 0\ ,
\end{equation}
where $r\in \frac{1}{w}\,\mbb{Z}$. The ground state of the spectrally flowed sector is (for odd $w$) \cite{Dei:2020zui}
\begin{equation}
    \Phi^w = [\ket{m_1^0,m_2^0}]^{\sigma^w}\otimes \ket{0}_\mrm{NS}\ ,\quad m_1^0= -\tfrac{(w-1)^2}{8\,w}\ ,\,\,m_2^0 = -\tfrac{(w+1)^2}{8\,w}\ .
\end{equation}
Applying $\mcl{K}^+_{-1/w}$ to $\Phi^w$ gives therefore 
\begin{align}
    \underset{C(0)}{\oint}\frac{dz}{2\pi i}\,\, K^+(z)\,\left(\,\tfrac{1}{z}\,\left[\Ket{m_1^0+\tfrac{1}{2\,w},m_2^0+\tfrac{1}{2\,w}}\right]^{\sigma^w}\otimes \ket{0}_\mrm{NS} + \mcl{O}(1)\right) &= \nonumber \\
    &\hspace{-4.6cm}=\left[\Ket{m_1^0+\tfrac{1}{2\,w},m_2^0+\tfrac{1}{2\,w}}\right]^{\sigma^w}\otimes K^+_{-1}\,\ket{0}_\mrm{NS}\ .
\end{align}
For  $\mcl{K}^+_{-3/w}$, the calculation works similarly, except that the $\gamma^{-3/w}$ factor now produces poles of order three and smaller. However, essentially by the same argument as in Section~\ref{sec:bps_states}, only the highest order pole contributes. Iterating the argument in this manner, we thus conclude that 
\begin{subequations}
\begin{align}
    &\mcl{K}^+_{-1}\cdots \mcl{K}^+_{-3/w}\,\mcl{K}^+_{-1/w}\,\Phi^w = [\ket{1/2,0}]^{\sigma^w}\otimes K^+_{-w}\cdots K^+_{-3}\,K^+_{-1}\,\ket{0}_\mrm{NS}\ ,\\
    &\mcl{K}^+_{-(w-2)/w}\cdots \mcl{K}^+_{-3/w}\,\mcl{K}^+_{-1/w}\,\Phi^w = [\ket{0,-1/2}]^{\sigma^w}\otimes K^+_{-(w-2)}\cdots K^+_{-3}\,K^+_{-1}\,\ket{0}_\mrm{NS}\ ,
\end{align}
\end{subequations}
which reproduces exactly the states we determined before, see eq.~(\ref{eq:dual_bps_states_full}). The calculation for even $w$ works similarly.

\section{Conclusion}\label{sec:conclusion}

In this paper we have shown that the BPS correlators of the symmetric orbifold are reproduced correctly from the string worldsheet perspective, i.e.\ from string theory on $\text{AdS}_3\times {\rm S}^3\times \mbb{T}^4$ with $k=1$ units of NS-NS flux. The key step in our argument was to observe that the worldsheet states that are dual to the BPS states of the symmetric orbifold can essentially be identified with the lift of the BPS states under the covering map; then the matching of the correlators is manifest, and we have also checked it explicitly for some simple 3-point functions. The proposed relation between spacetime and worldsheet states obviously fits very nicely with the general idea that the worldsheet plays the role of the covering surface in this duality. 

Given the relative simplicity of this identification it suggests that the result should be true more generally, and we have shown that, at least formally, the argument also works for any $\mfr{su}(2)_1$ descendant. In that case, the dual worldsheet state can be described by applying the corresponding DDF operators of \cite{DelGiudice:1971yjh,Giveon:1998ns,Eberhardt:2019qcl} to the twisted sector ground state. We only spelled this out for the $\mfr{su}(2)_1$ descendants, but one should expect this argument to work even more generally, and it would be interesting to work this out. The key technical problem will be to really make sense of the Wakimoto field $\gamma(z)$  which seems to behave like the covering map in all correlators.

\section*{Acknowledgements} We thank Lorenz Eberhardt, Rajesh Gopakumar, Bob Knighton and Kiarash Naderi for useful discussions and comments on a draft version of this paper. This paper is based on the Master thesis of BN. The work of the group is supported by the NCCR SwissMAP which is funded by the Swiss National Science Foundation.

\appendix

\section{Calculation of Dual BPS Three-Point Functions}\label{sec:app_calculations}

In this appendix we calculate the fermionic part of three-point functions of BPS states on the worldsheet explicitly. As we shall see, combining this with the symplectic boson part and considering ratios of correlators (that do not depend on the undetermined overall cofficient) we can reproduce the results of \cite{Lunin:2001pw} for many cases; we also conjecture a general form of the correlator which reproduces the ratios of their conjectured general result. 

Recall that the symplectic boson part of the three-point correlator is described by (\ref{eq:symplectic_boson_correlator}). For three punctures, a unique branched covering exists for any choice of $w_i$ and $x_i$ provided that the $w_i$ satisfy
\begin{equation}
    w_1+w_2+w_3\in 2\,\mbb{Z} + 1\ ,\quad w_{i}+w_{i+1} \geq w_{i+2} + 1\ .
\end{equation}
The coefficients $a_i$ for the corresponding covering map were found in \cite{Lunin:2000yv}, and are explicitly given by
\begin{equation}\label{eq:cov_map_coeff}
    a_i = \frac{d!\,(d-w_{i+1})!\,(d-w_{i+2})!}{w_i!\,(w_i-1)!\,(d-w_i)!}\ ,
\end{equation}
where $d = \frac{1}{2}\,(w_1 + w_2 + w_3 - 1)$ is the degree of the branched covering.

The fermionic part of the correlator is a pure $\mfr{su}(2)_1$ correlator after the $W$ and $\mcl{Q}$ fields have been contracted with one another. Moreover, by charge conservation, the dependence on the choice of $\mfr{su}(2)$ descendants is described by the 3j symbols, i.e.\ by the Clebsch-Gordan coefficients of the three $\mfr{su}(2)$ representations fusing into the singlet,
\begin{equation}
\ket{0,0} = \sum_{m_1,\,m_2,\,m_3}\begin{pmatrix}
j_1 & j_2 & j_3 \\
m_1 & m_2 & m_3
\end{pmatrix}\,\ket{j_1,\,m_1}\otimes \ket{j_2,\,m_2}\otimes\ket{j_3,\,m_3}\ ,
\end{equation}
where $\ket{j,m}$ is the state with Casimir $j(j+1)$ and $J^3_0$ eigenvalue $m$. For the following it is convenient to take the fermionic part of the three dual BPS states (\ref{eq:dual_bps_states_full}) as $\ket{j,-j}$, $\ket{k,k}$, and $\ket{l,j-k}$. Written in terms of the free fermions, the highest weight states are 
\begin{equation}
    \ket{k,k} = \psi^+_{-(2k-1)/2}\chi^+_{-(2k-1)/2}\cdots \psi^+_{-1/2}\chi^+_{-1/2}\ket{0}_\mrm{NS}\ .
\end{equation}
To find the fermionic part, we then need to calculate the proportionality constant $c^\mrm{ferm}_{jkl}$ in
\begin{align}\label{eq:ferm_corr}
\begin{split}
&\left\langle V(\ket{j,\,-j},z_1)\,V(\ket{k,\,k},z_2)\,V(\ket{l,\,j-k},z_3) \right\rangle =\\[0.3\baselineskip]
&\hspace{2.75cm}=\frac{c^\mrm{ferm}_{jkl} \,\begin{pmatrix}
l & k & j\\
j-k & k & -j
\end{pmatrix}}{(z_1-z_2)^{j^2+k^2-l^2}\,(z_2-z_3)^{k^2+l^2-j^2}\,(z_3-z_1)^{l^2+j^2-k^2}}\ .
\end{split}
\end{align}
These fields are holomorphic and primary, and one may use the techniques of e.g.\ 
\cite{Goddard} to write this constant in terms of the matrix element,
\begin{align}\label{eq:matrix_element_to_calculate}
c^\mrm{ferm}_{jkl}\,\begin{pmatrix}
l & j & k\\
j-k & -j & k
\end{pmatrix} &= \bra{j,\,j}V(\ket{k,\,k})_{l^2-j^2}\ket{l,\,j-k}\nonumber\\
&= \sqrt{\frac{(l+j-k)!}{(l-j+k)!\,(2l)!}}\,\bra{j,\,j}V(\ket{k,\,k})_{l^2-j^2}(K_0^-)^{l-j+k}\ket{l,l}\ .
\end{align}
The matrix element can be brought into a simple form by writing the vertex operators in an antisymmetrised form
\begin{equation}
    V(\ket{k,k})_m = \sum_{r_i,\,s_j}P_k(r_i,\,s_j)
    \,\psi^+_{r_k}\chi^+_{s_k}\cdots \psi^+_{r_1}\chi^+_{s_1}\,\delta_{\sum r_i+s_i,\,m}\ ,
\end{equation}
with
\begin{equation}
    P_k(r_i,\,s_j) = \frac{1}{(1!\,2!\cdots (k-1)!\,k!)^2}\prod_{a < b}(r_a-r_b)\,(s_a-s_b)\ .
\end{equation}
Using then the simple commutation relations of the free fermions, one finds the expression
\begin{align}
\begin{split}
    \bra{j,j}V(\ket{k,k})_{l^2-j^2}(K_0^-)^n\ket{l,l} &= \sum^{(2l-1)/2}_{r_i,\,s_j = -(2j-1)/2}\,\,\,\frac{(-1)^k\,n!\,(k!)^2}{(2k-n)!}\, P_k(r_i,s_j)\\
    &\hspace{3cm}\times \delta_{r_k,\,(2j+1)/2}\cdots \delta_{r_{2k-n+1},\,(2l-1)/2}\\
    &\hspace{3cm}\times \delta_{s_k,\,(2j+1)/2}\cdots \delta_{s_{2k-n+1},\,(2l-1)/2}\\
    &\hspace{3cm}\times \delta_{r_{2k-n}+s_{2k-n},\,0}\cdots \delta_{r_1+s_1,\,0}\ .
\end{split}
\end{align}
This expression can be evaluated with \texttt{Mathematica}, and we find experimentally for $k\leq 6$ that
\begin{align}\label{eq:bps_corr_end_result}
    \bra{j,j}V(\ket{k,k})_{l^2-j^2}(K_0^-)^n\ket{l,l} &= \nonumber \\[0.3\baselineskip]
    &\hspace{-2cm} = (-1)^k\,c(k,n)\,\frac{(2j+2(n-k))!}{(2j+2(n-k)-n)!}\cdots \frac{(2j+n-1)!}{(2j-1)!}\ ,
\end{align}
where $n=l-j+k$ and the coefficients $c(k,n)$ are fixed by $c(k,2k) = (2k)!$ and
\begin{equation}\label{eq:coeff_rec_rel}
   \frac{c(k,n+1)}{c(k,n)} = \frac{(n+1)!}{(2k-(n+1))!}\ .
\end{equation}
This reproduces the orbifold result provided that
\begin{equation}\label{eq:matching_fractions}
    \frac{c^\mrm{ferm}_{jk(l+1)}}{c^\mrm{ferm}_{jkl}} = a_3\,\frac{\hat{C}^{+-+}_{w_1,w_2,w_3}}{\hat{C}^{+--}_{w_1,w_2,w_3}}\ ,
\end{equation}
where the $\hat{C}^{1_11_21_3}_{w_1,w_2,w_3}$ are the coefficients found in \cite[eq.~(6.39)]{Lunin:2001pw}, and we consider $w_1 = 2j-1$, $w_2 = 2k+1$, and $w_3 = 2l+1$ for definiteness.\footnote{We consider ratios of correlators here to remove the overall constant unfixed by the incidence relation. The relative factor of $a_3$ arises from the orbifold because of eq.~(\ref{eq:cov_map_coeff_exponent}). For $w_2 = 2k-1$, the extremal value $l=j+k$ is not allowed, i.e.\ no covering map exists and the bosonic correlator vanishes.} Explicitly, the worldsheet ratios are given by
\begin{align}
\begin{split}
        \frac{c^\mrm{ferm}_{jk(l+1)}}{c^\mrm{ferm}_{jkl}} =\frac{c(k,n+1)}{c(k,n)}\,& \frac{(2j+n)!}{(2j+2(n-k)+1)!}\\
        &\hspace{-1.3cm}\times \sqrt{ \frac{(2j+n+2)\,(2j+2(n-k)-n+1)!\,(2j+2(n-k)-n)!}{(n+1)\,(2k-n)\,(2j+2(n-k)+2)!\,(2j+2(n-k))!} }\ .
\end{split}
\end{align}
Plugging this into eq.~(\ref{eq:matching_fractions}) we find that the identity is satisfied if the $c(k,n)$ satisfy (\ref{eq:coeff_rec_rel}).

\bibliographystyle{JHEP}

\end{document}